\documentclass[3p,12pt]{elsarticle}

\usepackage{natbib}

\usepackage{comment}

\usepackage{hyperref}

\usepackage{array}
\usepackage{subcaption}

\usepackage{siunitx}
\DeclareSIUnit{\fC}{\femto\coulomb}
\DeclareSIUnit{\mV}{\milli\volt} 

\newcommand{\dNdx}{\ensuremath{dN/dx}}
\newcommand{\dEdx}{\ensuremath{dE/dx}}

\usepackage{amssymb}

\usepackage{amsmath}

\journal{Nuclear Instruments and Methods in Physics Research - section A}

\begin{document}

\begin{frontmatter}

\title{A 24-Channel Ultra-Low-Noise Preamplifier for dN/dx Measurements with Drift Tube Detectors}

\author[a]{Jiajin Ge}
\author[a]{Chihao Li}
\author[a]{Can Suslu}
\author[a]{Yuxiang Guo}
\author[a]{Emmett Salzer}
\author[a]{Tiesheng Dai}
\author[a]{Jianming Qian}
\author[a]{Bing Zhou}
\author[a]{Junjie Zhu\corref{cor1}}

\cortext[cor1]{Corresponding author. Email Address: junjie@umich.edu} 

\affiliation[a]{organization={Department of Physics},
            addressline={University of Michigan}, 
            city={Ann Arbor},
            state={MI},
            postcode={48109}, 
            country={USA}}
            
\begin{abstract}
Cluster counting (\dNdx) is a promising method to enhance particle identification for gaseous detectors, especially in next-generation collider experiments like the FCC-ee, where good $\pi/K$ separation over a broad momentum range is essential. However, its implementation in large-scale systems has been limited by the challenging requirements for high-resolution signal amplification and readout. This paper presents a 24-channel ultra-low-noise preamplifier board designed for drift tube detectors to enable \dNdx\ measurements. The three-stage amplification topology employs SiGe transistors and integrates dedicated noise-minimization techniques, achieving a charge gain of \SI{21.11}{ \mV/\fC} from \SI{0.3}{\femto\coulomb} to \SI{50}{\femto\coulomb}, a bandwidth of \SI{542}{\mega\hertz}, and a voltage gain of \SI{47.8}{dB}. The measured voltage noise density is \SI{0.35}{\nano\volt\per\sqrt{\hertz}}, surpassing most of the state-of-the-art preamplifiers for gaseous and silicon detectors. Validation tests conducted on the sMDT chambers at the CERN Proton Synchrotron test beam facility demonstrate that the proposed design meets the stringent preamplifier requirements for implementing the \dNdx\ method in drift-tube detector systems, achieving an equivalent noise charge of \SI{0.14}{\femto\coulomb} and a signal-to-noise ratio of 73 when operated with a He:iC\textsubscript{4}H\textsubscript{10} (90:10) gas mixture. The design also shows promise for broader application in other gaseous or semiconductor detectors.

\end{abstract}

\begin{graphicalabstract}
\end{graphicalabstract}

\begin{highlights}
\item A 24-channel low-noise preamplifier designed for drift-tube detector readout with cluster-counting capability (\dNdx).
\item The circuit achieves  \SI{0.35}{\nano\volt\per\sqrt{\hertz}} noise density with \SI{542}{\mega\hertz} bandwidth and 47.8 dB voltage gain, surpassing most existing gaseous and silicon preamplifiers.
\item Validated on sMDT chambers in a \SI{1}{GeV} hadron beam with a He:iC\textsubscript{4}H\textsubscript{10} (90:10) gas mixture, the proposed preamplifier meets the stringent front-end requirements for the \dNdx\ method in drift-tube detectors, achieving an ENC of \SI{0.14}{\femto\coulomb} and an SNR of 73.

\end{highlights}


\end{frontmatter}

\section{Introduction}
\label{sec1}
Particle identification (PID) is a cornerstone of particle physics, playing a critical role in advancing the field --- from probing rare processes to searching for new phenomena. In gaseous detectors, the traditional method for PID relies on measuring a particle’s energy loss per unit length (\dEdx) as it traverses a medium~\cite{Bethe:1930ku, Allison:1980vw, ParticleDataGroup:2024cfk, Rolandi:2008ujz, Sauli:2014cyf}. This technique leverages the dependence of energy loss on the particle’s speed and charge. However, its discriminating power diminishes at high momenta, where energy loss rates for different particle species become nearly indistinguishable. Furthermore, intrinsic fluctuations in energy deposition and variations in the gas gain limit its resolution.

A promising alternative is the cluster-counting approach (\dNdx)~\cite{Rolandi:2008ujz}, which determines PID by counting the number of primary ionization clusters per unit length. When many clusters are produced along a particle’s path, the uncertainty in cluster counting approximates a Gaussian distribution, scaling as $1/\sqrt{N_{\text{cluster}}}$, where $N_{\text{cluster}}$ is the number of detected primary ionization clusters. This method leads to significantly enhanced PID performance, especially for particles with multi-GeV momenta. For instance, cluster counting can nearly double the significance of the $\pi/K$ separation relative to the \dEdx\ method~\cite{Vavra:2000vag, Elmetenawee_2025}. This level of improvement is particularly crucial for proposed FCC-ee experiments, which require the inner tracker system to achieve better than $3\sigma$ separation between pions and kaons across a broad momentum range, from $\mathcal{O}$(100 MeV) to $\mathcal{O}$(40 GeV).

While promising, the \dNdx\ method has not yet been implemented in operational experiments and remains an active area of research and development. The technique poses significant challenges, most notably the readout of single-ionization electron signals and the reliable identification of closely-spaced ionization clusters. Achieving this requires fast, highly sensitive electronics capable of processing high-resolution detector signals in real time. Although the concept was introduced more than 50 years ago, no large-scale experiment has implemented the \dNdx\ technique to date.

The \dNdx\ method relies on the detection of signals created by single electrons and sensitive front-end (FE) electronics are needed. This paper describes the design and performance of an FE board that amplifies single-ionization electron signals in gaseous chambers. 

\section{sMDT Detector and Front-End Requirements}
\label{sec_smdt}
The small-diameter Monitored Drift Tube (sMDT) detectors are used \cite{ATLAS-TDR-026}. For FCC-ee experiments, gaseous detectors such as drift chambers or straw tubes are proposed due to their low material budget, which enables improved momentum resolution, especially at low momenta. The performance of the \dNdx\ method depends primarily on the electric field configuration, the gas mixture, and the gas pressure. However, it is independent of the material of the chamber wall. The insights gained from the sMDT detector are thus broadly applicable across similar detector technologies.

\subsection{The sMDT Detector}
\label{smdt_subsec1}
The sMDT detector consists of multiple layers of aluminum tubes with a diameter of \SI{15}{\milli\meter}. At the center of each tube is a tungsten-rhenium anode wire with a diameter of \SI{50}{\micro\meter} and a resistance of about \SI{44}{\ohm/\meter} \cite{ATLAS-TDR-026}. During operation, the tubes are filled with a controlled gas mixture---typically Ar:CO\textsubscript{2} or He:iC\textsubscript{4}H\textsubscript{10} at a specified pressure. A positive high voltage (HV) is applied to the anode wire, while the tube wall is held at ground potential. 

When a charged particle traverses the active volume, it ionizes the gas and produces electron-ion pairs. While most of these primary electrons have low energies, some can receive enough energy to become delta electrons, which then initiate further ionizations and produce additional secondary electrons. Both primary and secondary electrons drift to the anode wire. As they approach the wire, these electrons gain sufficient energy to trigger additional ionizations, leading to an avalanche of electrons around the wire. The induced current pulse forms the measurable signal on the wire. These signals manifest as short pulses, with their shape and amplitude determined by the number of electrons involved and stochastic variations in the avalanche process.

\begin{figure}[htbp]

\centering
\includegraphics[width=1\textwidth]{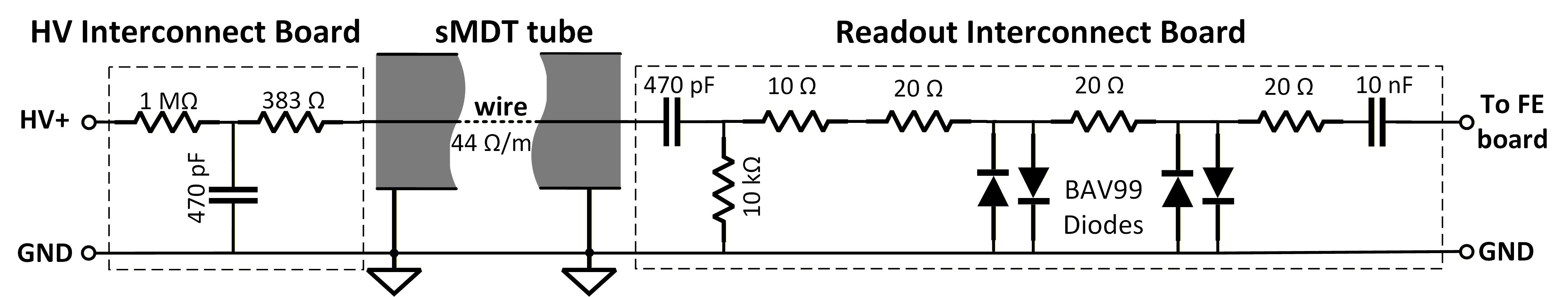}
\caption{Schematic of ATLAS sMDT on-chamber electronics: HV and readout interconnect boards.}\label{sMDTandHedgehogCircuit}
\end{figure}

Signals are read out from one end of the anode wire, while the other end is connected to the HV supply. Both the HV side and the readout side are equipped with dedicated electronics boards housed in a Faraday cage: the HV interconnect ("hedgehog") board for HV distribution and the readout interconnect board for signal AC coupling with two-stage protection circuitry \cite{ATLAS-TDR-026, Arai_2008}. A schematic of the sMDT tube and interconnect boards assembly is shown in Figure~\ref{sMDTandHedgehogCircuit}. The readout interconnect board has a series resistance of \SI{70}{\ohm}, which matches the \SI{70}{\ohm} resistance of the \SI{1.6}{m} long anode wire inside the ATLAS sMDT tube. Each readout interconnect board connects to 24 sMDT tubes. The sMDT FE board, which is mounted on the readout interconnect board, integrates three 8-channel amplifier-shaper-discriminator (ASD) chips for signal conditioning \cite{Penski_2024}. The discriminated outputs are subsequently processed by a 24-channel time-to-digital converter (TDC). The TDC records both the time of the first arrival signal and the pulse width. The pulse width is determined by a Wilkinson analog-to-digital converter, which measures the time required for the integrated signal charge over the initial $\sim 20$ ns to discharge from a capacitor \cite{ATLAS-TDR-026}. To prevent multiple threshold crossings of a single signal, the ASD ASIC is programmed with a dead time of $\sim 1 \mu s$. After detecting the earliest arrival signal, the ASD ignores additional crossings and no further time measurements are recorded during this dead time interval. 

In the ATLAS MDT system, it is sufficient to record only the time of the earliest arriving signal and a coarse estimation of signal amplitude. However, the \dNdx\ method requires capturing complete waveforms over several hundred nanoseconds with time resolution better than 1 ns. This can be achieved through fast waveform-digitization or analog-timing-feature extraction, both of which demand sophisticated electronics designed to process high-frequency signals. Accurate cluster counting may require a delicate balance between two conflicting timing requirements: (1) fine time resolution to minimize signal overlap from primary electrons produced in separate ionization events; and (2) coarser time resolution to have secondary electrons from the same primary cluster arrive sufficiently close in time to avoid being misidentified as separate clusters.

\subsection{Requirements for the preamplifier}
\label{smdt_subsec2}
For \dNdx\ measurements in drift‑tube detectors, a gas gain on the order of \(10^{5}\) is commonly applied \cite{Elmetenawee_2025}. Under such a gain, the total charge induced by a single electron can be estimated as:
\begin{equation}
Q_{\text{single}} = e \times G \times \eta = 1.6 \times 10^{-19}\,\text{C} \times 10^{5} \times 0.1 = 1.6\,\text{fC}
\label{eq:single_electron_induced_charge}
\end{equation}
where \(e\) denotes the elementary charge, \(G\) the gas gain, and \(\eta\) a multiplicative factor (taken here as 0.1) accounting for the fact that avalanche multiplication is confined to a region typically within \SIrange{20}{100}{\micro\meter} from the anode wire surface. In practice, this charge may be further reduced by a factor of $\sim 2$ due to signal reflection, dissipation, and attenuation resulting from the resistance of the wire. In this case, the avalanche output of a signal induced by a single electron is about \SI{0.8}{\femto\coulomb}. Given that a minimum signal‑to‑noise ratio (SNR) of 3 is required, the equivalent noise charge (ENC) of the preamplifier must therefore be better than \SI{0.27}{\femto\coulomb}.

The leading edge time of a typical sMDT raw signal is approximately \SI{1}{\nano\second} \cite{ PANAREO2023167822}. To amplify the signal without distortion, the bandwidth of the preamplifier must exceed the knee frequency \( f_{\mathrm{knee}} \) of the signal, which is estimated as:
\begin{equation}
f_{\mathrm{knee}} \approx \frac{0.35}{t_{\mathrm{rise}}}
\label{eq:knee_freq}
\end{equation}
Accordingly, the \SI{-3}{dB} bandwidth of the preamplifier should be greater than \SI{350}{\mega\hertz}.

\section{Electronic Design}
\label{sec_design}
A preamplifier board is designed to amplify the electron signals for our \dNdx\ R\&D studies with the sMDT detector. To ensure system compatibility, the preamplifier board is designed to handle 24 channels and maintains full electrical and mechanical compatibility with the ATLAS sMDT readout interconnect board. 

\begin{figure}[htbp]
\centering
\includegraphics[width=1\textwidth]{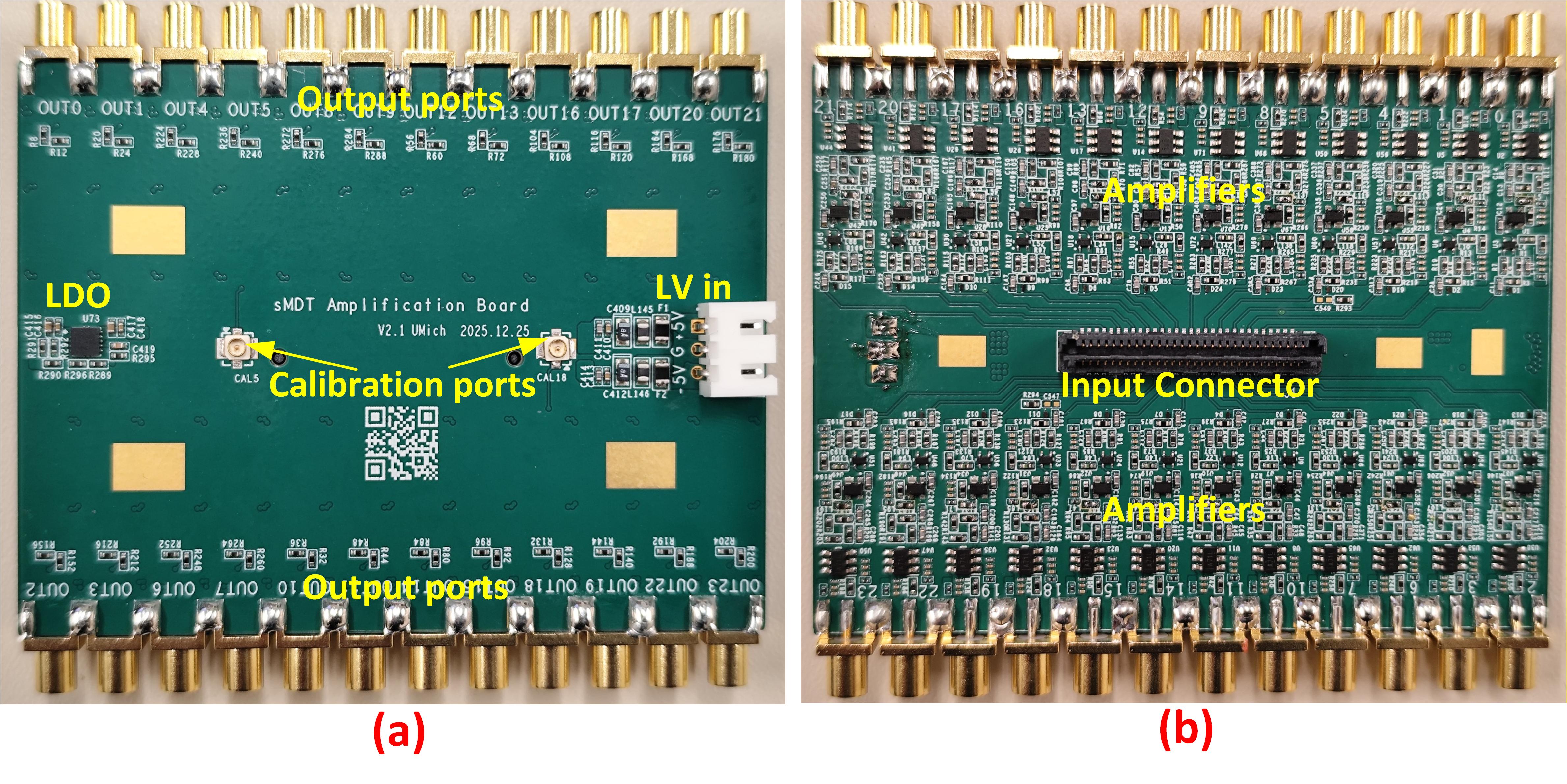}
\caption{Top (a) and bottom (b) views of the preamplifier board.}\label{board_picture}
\end{figure}

Figure~\ref{board_picture} illustrates the top and bottom views of the preamplifier board. The printed circuit board (PCB) has four layers, with a size of \SI{8.1}{\centi\meter} $\times$ \SI{6.1}{\centi\meter} to fit within the compact structure of the sMDT chambers. The board plugs into the sMDT readout interconnect board via a Samtec high-speed hermaphroditic connector (LSEM-130-06.0-F-DV-A-N-K-FR). The low-voltage (LV) power input and Low-Dropout regulator (LDO) are located on the top layer. Additionally, two calibration ports are provided, connected to channels 5 and 18, for tests with a signal generator and gain calibration.
Amplifier circuits for 24 channels are located on the bottom layer. 

\subsection{Implementation of the three-stage amplifier}
\label{design_subsec1}

\begin{figure}[htbp]
\centering
\includegraphics[width=1\textwidth]{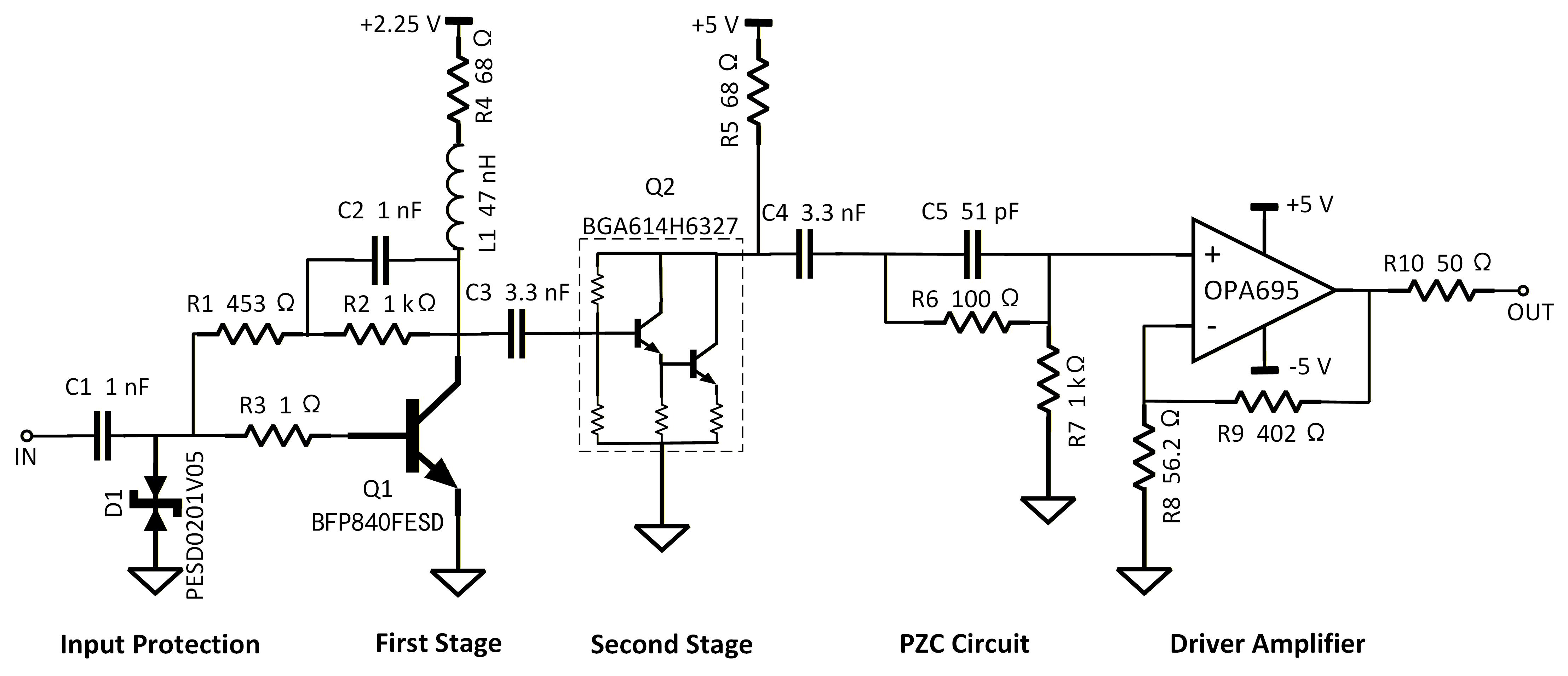}
\caption{Schematic of the three-stage amplifier for a single channel.}\label{Circuit_Schematic}
\end{figure}

To achieve an amplification factor of several hundred while maintaining a broad bandwidth, a multi-stage architecture with cascaded high-bandwidth amplifier circuits is typically adopted. In this design, we employ a three-stage amplification scheme, as shown in Figure~\ref{Circuit_Schematic}. Each preamplifier channel contains a protection circuit, three amplifier stages, and a Pole-Zero Cancellation (PZC) circuit.

The input signal from the sMDT readout interconnect board is AC coupled through a \SI{1}{\nF} capacitor ($C_1$). A Polymer Electrostatic Discharge (PESD) bi-directional protection diode (PESD0201V05) is used, featuring ultra-low capacitance ($C_j = \SI{0.05}{\pF}$) and low reverse leakage current (\SI{0.01}{\micro\ampere}). The design of the first stage is critical for achieving low noise in amplifier circuits. Building on our previous work \cite{GeLGADAmp}, we choose a Hetero-junction Bipolar Transistor (HBT) RF SiGe transistor (BFP840FESD), which has a transition frequency of \SI{85}{\GHz}, to form a common emitter amplifier. The collector current of the HBT is configured at approximately \SI{10}{\milli\ampere} via a resistive network formed by $R_1$, $R_2$, and $R_4$. This configuration results in a voltage gain of $\sim$\SI{18}{\dB} and a bandwidth up to \SI{1.9}{\GHz}, as specified in the manual \cite{BFP840FESD}. An isolation resistor ($R_3$) is placed near the base of $Q_1$ to suppress potential resonance caused by trace inductance and parasitic capacitance, effectively minimizing input noise. The BFP840FESD transistor also features integrated ESD protection rated at \SI{1.5}{\kV}, further enhancing robustness.

In our previous design for the Low‑Gain Avalanche Detector (LGAD) \cite{GeLGADAmp}, we utilized a high‑speed operational amplifier in the second stage to ensure a wide linear dynamic range. However, the gas avalanche process in the sMDT is relatively slow, with signal rise times ranging from several nanoseconds to tens of nanoseconds. As a result, the sMDT output signal amplitude is nearly an order of magnitude lower than that of the LGAD \cite{Rolandi:2008ujz, DubbertMDT, MoffatLGAD}. We adopt a transistor amplifier circuit for the second stage, reducing both noise and overall cost. Inspired by the work of Hoarau, et al. (2021) on pulse amplifier for diamond particle detectors, we use a Darlington-configured integrated SiGe broadband amplifier (BGA614) in the second stage to provide a voltage gain of $\sim$~\SI{10}{\dB} \cite{HoarauCVDAmp, BGA614}. The BGA614 offers a wider \SI{-3}{\dB} bandwidth of \SI{2.4}{\GHz} and a flatter frequency response compared to the BGA427 used in Ref. \cite{HoarauCVDAmp}, resolving the non-flat gain issue identified in their work.

To suppress the pulse undershoot, a PZC network is implemented using $C_5$, $R_6$, and $R_7$. The transfer function of the network can be expressed as follows:
\begin{equation}
H(s) = \frac{V_o(s)}{V_i(s)} = \frac{s + \frac{1}{\tau_1}}{s + \frac{1}{\tau_2}}, \quad \text{with} \, \tau_1 = R_6C_5, \, \tau_2  = \frac {R_6R_7}{R_6+R_7}C_5
\label{eq:transfer_function}
\end{equation}
This circuit achieves pole-zero cancellation (the zero at 1/$\tau_1$ cancels a pre-existing pole)\cite{BoieTailCancel}. In practice, the component values are determined to match the decay characteristics of the detector signal and then further fine-tuned experimentally to minimize undershoot and achieve optimal pulse shaping. In our design, $\tau_1$ is around \SI{5}{\ns}.

A third-stage amplifier uses an OPA695 operational amplifier to boost the signal gain, extend the output dynamic range, and strengthen the load-driving capability. The stage provides a gain of \SI{18}{\dB}, bringing the total theoretical gain of the entire three-stage amplifier to $\sim$\SI{46}{\dB}. The OPA695 limits the bandwidth to \SI{600}{\MHz} (adequate for this design) to suppress noise. Further upgrades of the third stage may be incorporated in the future to achieve even greater performance as needed. For example, the OPA695 can be easily replaced with higher-performance amplifiers (such as OPA847 or LMH6629) on the same PCB if wider bandwidth or higher gain is needed. For example, if LMH6629 is used, the bandwidth will increase to \SI{900}{\MHz} and the gain will increase by a factor of 1.25 \cite{ti2022lmh6629}.

\subsection{Approaches to noise reduction}
\label{design_subsec2}
For a multi-stage low-noise amplifier, the overall system noise figure is primarily determined by the first stage, as described by Friis noise formula \cite{Spieler2005}:
\begin{equation}
e_{n}^2 = e_{n,1}^2 + \frac{e_{n,2}^2}{A_1^2} + \frac{e_{n,3}^2}{(A_1 A_2)^2} + \dots
\label{eq: Friis_noise_formula}
\end{equation}
where \( e_n \) represents the equivalent input-referred voltage noise density of a cascaded system, \( e_{n,\text{i}} \) represents the input-referred voltage noise density of stage $i$, and \( A_i \) represents the voltage gain of stage $i$. Consequently, the primary design focus should be on optimizing the noise performance of the first stage. The BFP840FESD, with its low noise figure of \SI{0.55}{\dB} at \SI{0.45}{\GHz}, is well-suited for our application. By providing approximately tenfold gain in the first stage, the impact of noise from subsequent stages is effectively reduced, resulting in an improved overall SNR.

The ultra-low-noise LDO (TPS7A9401) provides a \SI{2.25}{\V} supply to the BFP840FESD amplifier. The root mean square (RMS) noise of the LDO is \SI{0.46}{\micro\volt}. To further suppress incoming noise from the power supply, an LC filter is placed at the board's power input. In addition, each amplifier’s power supply pin is locally decoupled using a ferrite bead and multiple capacitors of varying values positioned close to the pin.

The PCB layout is optimized by placing all amplifiers on the bottom layer. This allows the input signals to be routed directly from the connector to the amplifier inputs without using vias, minimizing parasitics and discontinuities in these sensitive paths. To further ensure signal integrity, a dense array of grounding vias is deployed along both sides of critical traces, between channels for isolation, and across unused areas of the board. This configuration establishes a low-impedance return path, thereby mitigating ground noise while suppressing electromagnetic interference (EMI) and crosstalk. For critical signal paths, 0402 surface-mount components are used to minimize parasitic effects. Signal traces within each channel are kept as short as possible to minimize inductive noise pickup. Despite the compact dimensions of the board, the minimum spacing between adjacent channels exceeds \SI{3}{mm}, and critical signal traces are separated by more than \SI{7}{mm} (40 times the trace width of \SI{7}{mil}). In addition, a continuous ground plane provides further isolation. Under these conditions, the crosstalk is expected to be negligible, consistent with typical signal integrity considerations \cite{bogatin2010signal}, and in agreement with the experimental observation that no noticeable crosstalk was observed.

\section{Test results}
\label{sec_test}
A series of measurements was carried out to evaluate the performance of the preamplifier board. Key electronic parameters, including gain, noise, linearity, and bandwidth, were characterized. In addition, the boards were integrated into the sMDT detectors to test the boards under realistic operating conditions. The test procedures and corresponding results are presented in the following subsections.

\subsection{Electronic performance tests}
\label{test_subsec1}
\begin{figure}[htbp]
\centering
\includegraphics[width=0.6\textwidth]{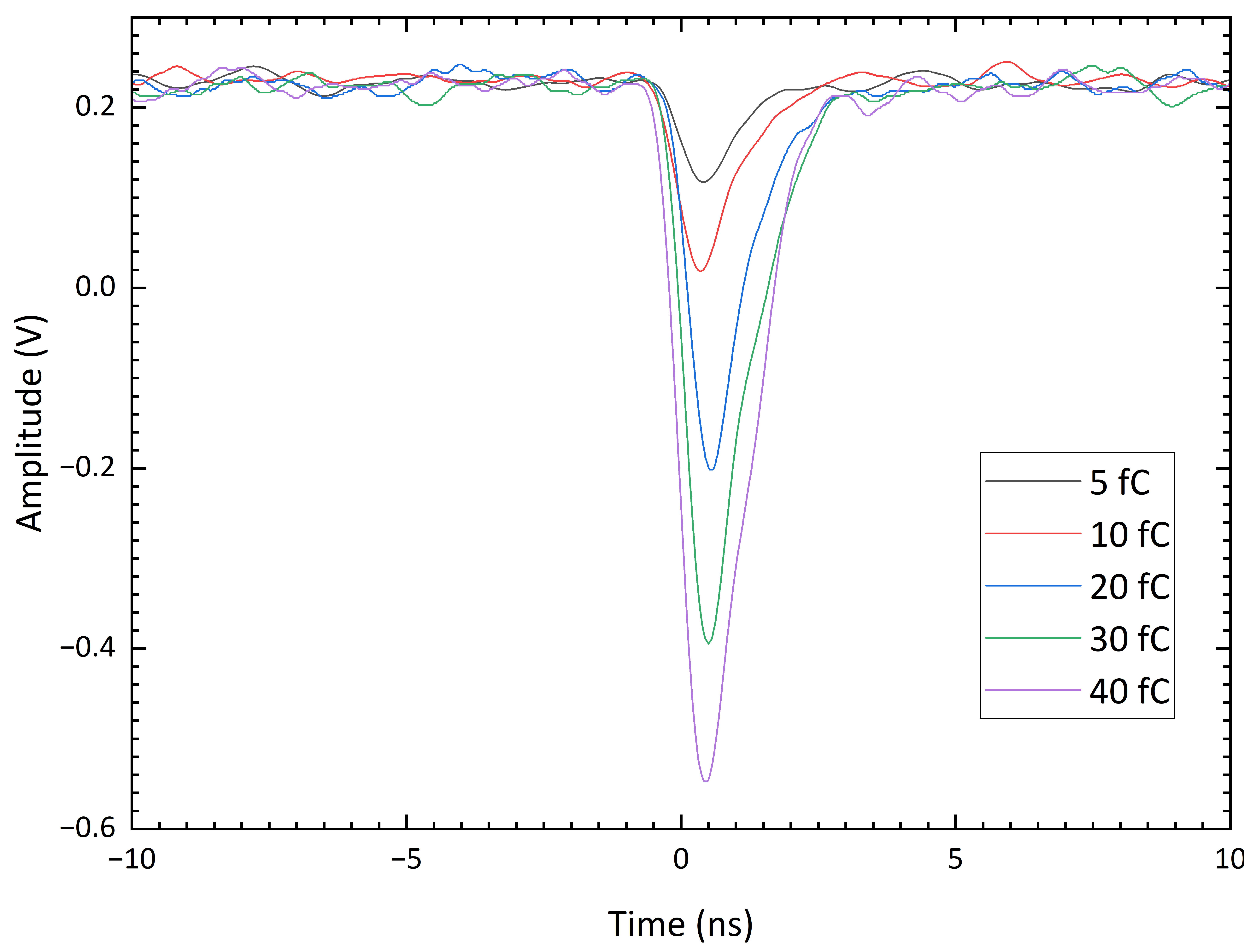}
\caption{Illustrative output waveforms under different injected charges.}\label{output_pulse}
\end{figure}

The charge gain and linearity of the amplifier were characterized using a fast pulse generator (ACTIVE PG-1072) with a typical rise/fall time of \SI{70}{\ps}. The pulse generator output was routed through a $-40$ dB or $-20$ dB attenuator before being connected to the calibration port of the preamplifier board. A \SI{50}{\ohm} termination resistor at the calibration input further halved the amplitude of the input pulse. The equivalent injected charge was calculated as follows:
\begin{equation}
Q_{\text{in}} = 0.01 \times 0.5 \times V_{\text{pulse}} \times C_{\text{cal}},
\label{eq:input_charge}
\end{equation}
where \( V_{\text{pulse}} \) is the amplitude of the input pulse, and \( C_{\text{cal}} \) is a \SI{1}{\pF} calibration capacitor (muRata GJM1555C1H1R0WB01) with a tolerance of \SI{0.05}{\pF} \cite{murata_gjm1555c1h1r0wb01d}, connected in series between the calibration port and the AC coupling capacitor (\( C_1 \)) of the amplifier. The output waveforms were captured and their average amplitudes were measured using a Rohde \& Schwarz RTA4004 oscilloscope. Representative output waveforms for different levels of injected charge are presented in Figure~\ref{output_pulse}. The measured rise time (10\% to 90\%) of the output pulse is approximately \SI{640}{\ps},  meeting the design requirements.

\begin{figure}[htbp]
\centering
\includegraphics[width=0.6\textwidth]{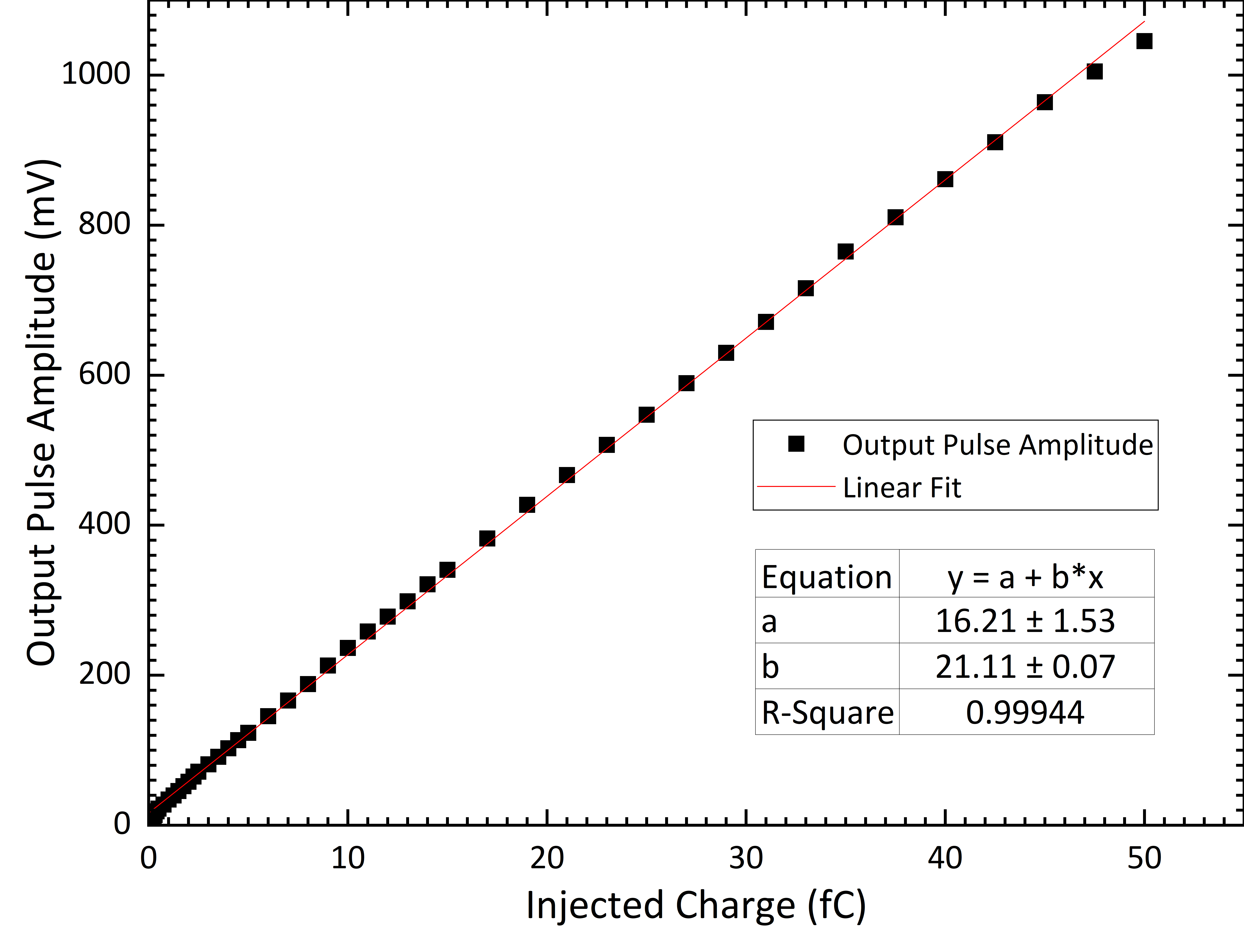}
\caption{Output pulse amplitude versus equivalent injected charge.}\label{charge_gain_plot}
\end{figure}

The output amplitude as a function of the equivalent injected charge is shown in Figure~\ref{charge_gain_plot}. The preamplifier has a measured charge gain of \SI{21.11}{ \mV/\fC}. Its effective dynamic range spans from \SI{0.3}{\fC} to \SI{50}{\fC}, defined as the range where the output amplitude exceeds 3.3 times the RMS noise level while maintaining linearity. Within this range, the probability of the peak noise surpassing 3.3 times the RMS value (relative to a zero baseline) is less than \SI{0.1}{\percent}. A linear fit to the data yields a \( R^2 \) value of 0.99944, confirming its excellent linearity.

\begin{figure}[htbp]
\centering
\includegraphics[width=0.8\textwidth]{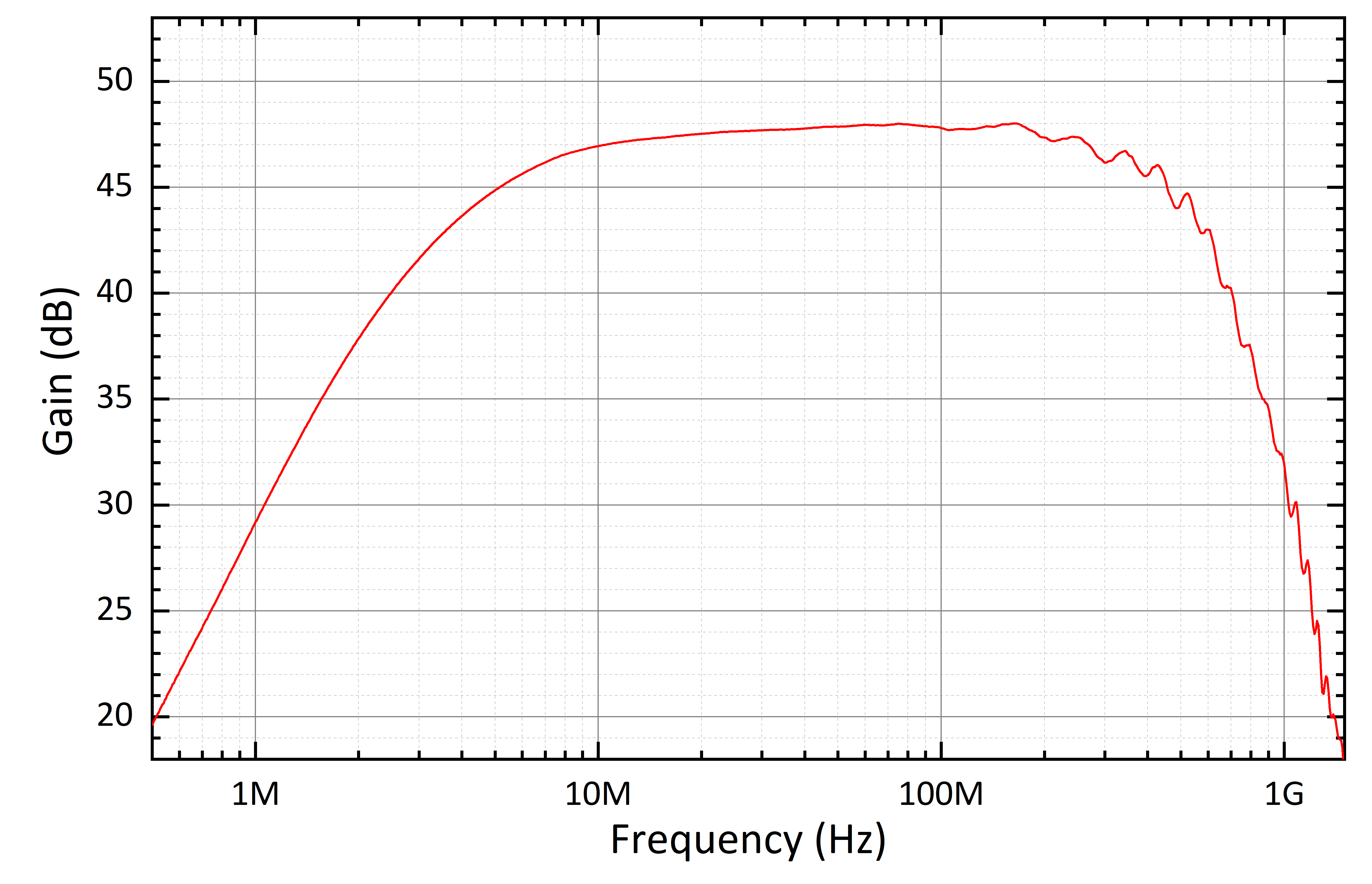}
\caption{Frequency response curve of a single preamplifier channel.}\label{bandwidth}
\end{figure}

The frequency response of the preamplifier was characterized using a network analyzer (Agilent E5061A). A swept-frequency signal from the analyzer's source port (Port $S_1$) was attenuated by \(-\SI{40}{\dB}\) to match the input range of the circuit board. Prior to measurement, both open- and through-circuit calibrations were conducted. As shown in Figure~\ref{bandwidth}, the resulting frequency response exhibits a bandwidth of approximately \SI{4}{\MHz} to \SI{542}{\MHz}, with a passband gain of $\sim$ \SI{47.8}{\dB} (245.5~V/V). The output shows a leading-edge transition time as fast as \SI{650}{\ps}, which is consistent with our test results. This implies that two successive electron clusters can be resolved if they are separated by more than $\sim 1$~ns, considering the noise. Such timing performance indicates that the preamplifier can distinguish closely spaced clusters, supporting its suitability for precise $dN/dx$ measurements in drift tube detectors.

\begin{figure}[htbp]
\centering
\includegraphics[width=0.6\textwidth]{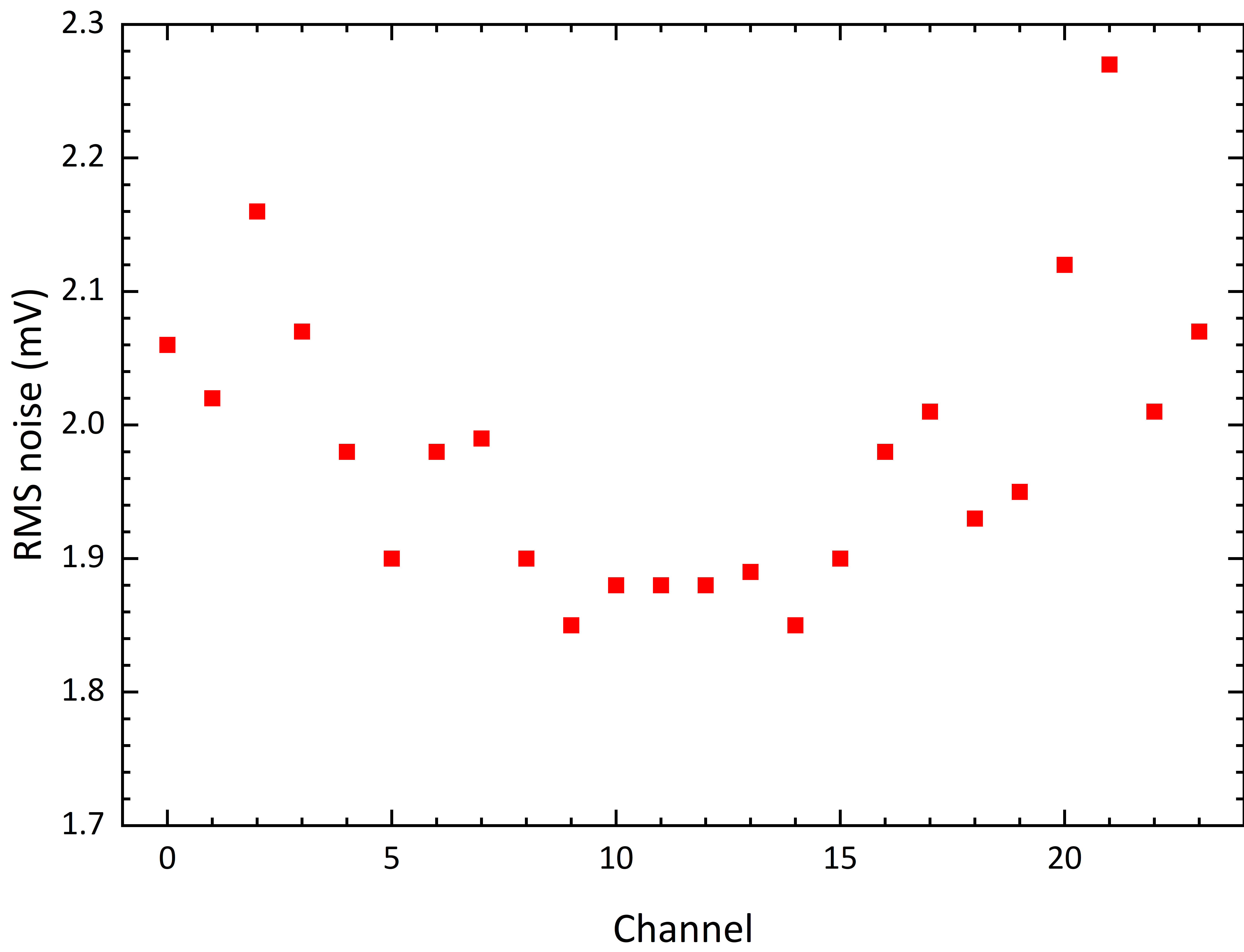}
\caption{Measured RMS noise for each channel.}
\label{noise_Results}
\end{figure}

The output RMS noise across all 24 channels was measured with inputs floating, as shown in Figure~\ref{noise_Results}. Although conventional noise measurements typically terminate the input with a matching resistor, the open-input condition was used here because it reflects the actual drift tube detector interface, which is high-impedance and DC-open. We acknowledge that this method may be more susceptible to environmental interference, potentially overestimating the true amplifier noise. Nevertheless, it remains the most relevant approach for evaluating noise performance under actual drift tube conditions. The average RMS noise is \( 1.98\,\text{mV}_{\text{rms}} \). Channels near the board edges have slightly higher noise due to longer input traces. The noise on one specific channel is noticeably higher because its input is located close to the power supply input. Using the measured output RMS noise \( V_{\text{out,rms}} = 1.98\,\text{mV} \), bandwidth \( BW \approx 542\,\text{MHz} \), and voltage gain \( G_v \approx 245.5 \), the input-referred voltage noise density \( e_n \) is calculated as \cite{Spieler2005}:
\begin{equation}
e_n = \frac{V_{\text{out,rms}}}{\sqrt{BW} \cdot G_v} \approx 0.35\,\text{nV}/\sqrt{\text{Hz}}
\label{eq:noise_density}
\end{equation}

This measured noise level is remarkably low for a 24-channel wide-band amplifier operating in the sub-GHz regime, attaining performance comparable to state-of-the-art, low-noise preamplifiers employed in high-energy physics, as compared in Table~\ref{tab:preamplifier_comparison}. For drift tube detectors with outer diameters ranging from \SI{5}{mm} to \SI{30}{mm}, the corresponding capacitance per unit length, calculated using the standard cylindrical capacitance formula, is approximately between \SI{8.7}{pF/m} and \SI{12.6}{pF/m}. In large-scale experiments, drift tube detectors are typically several meters long. Additionally, the readout interconnect board usually adds several picofarads of parasitic capacitance. Consequently, the total input capacitance of a drift tube detector is much larger than that of typical semiconductor detectors such as LGADs or CVD diamond detectors, whose capacitance is normally below \SI{5}{pF}. Remarkably, our amplifier maintains its low-noise performance despite this relatively large input capacitance of the drift tube detectors. Such low-noise characteristics enable the detection of weak signals down to the sub-fC level, directly reflecting a careful design that has successfully minimized all major noise contributions.

\begin{table}[htbp]
\centering
\footnotesize  
\caption{Performance comparison with existing low-noise preamplifiers.}
\label{tab:preamplifier_comparison}
\begin{tabular}{@{}lcccc@{}}
\hline
Amplifier & Channel & Gain (dB) & Bandwidth (MHz) & $e_n$ (nV/$\sqrt{\text{Hz}}$) \\
\hline\hline
WaveDAQ Drift tube Preamplifier~\cite{Ritt2015WaveDAQ} & 16 & 40 & 800 & 1.17 \\
Drift Tube VGA~\cite{PANAREO2023167822} & 1 & 38.43 & 927 & 2.30 \\
MEG II Drift Tube FE~\cite{CHIARELLO2016336}  & 8 & 20 & 884 & 6.73 \\
ATLAS sMDT ASD2~\cite{Penski_2024,ASD2_IEEESensor} & 8 & -- & 11 & 5 \\
CVD-diamond amplifier~\cite{HoarauCVDAmp} & 1 & 43 & 1000 & 0.63 \\
LGAD amplifier~\cite{GeLGADAmp} & 1 & 48.5 & 872 & 0.61 \\
LMH6629 operational amplifier~\cite{ti2022lmh6629} & 1 & 20 & 900 & 0.69 \\
\bfseries This work & \bfseries 24 & \bfseries 47.8 & \bfseries 542 & \bfseries 0.35 \\
\hline
\end{tabular}
\end{table}

\subsection{Tests on sMDT detector}
\label{test_subsec2}

\begin{figure}[htbp]
\centering
\includegraphics[width=0.8\textwidth]{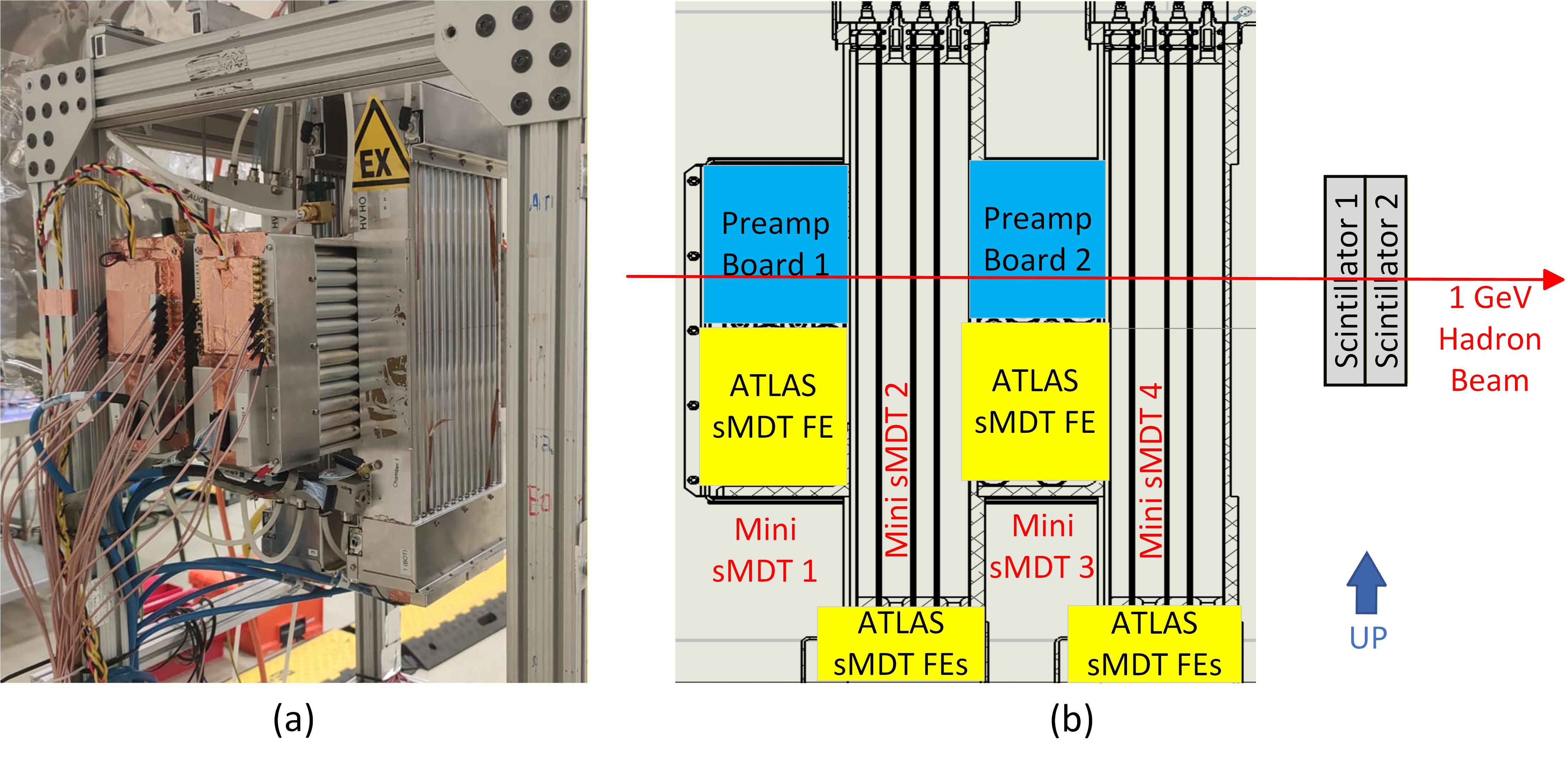}
\caption{(a) Photograph of the preamplifier boards mounted on the mini sMDT chambers; (b) Schematic of the mini sMDT test setup at the CERN PS T9 test beam facility.}\label{TestbeamSetup}
\end{figure}

The functionality of the preamplifier boards was further evaluated on the sMDT detectors in a test beam at the CERN Proton Synchrotron (PS) T9 facility. The experimental setup, shown in Figure~\ref{TestbeamSetup}, consisted of two horizontally placed mini sMDT chambers and two vertically placed mini sMDT chambers. The mini sMDT detectors share the same tube structure and detector geometry as the ATLAS sMDT, but feature a reduced number of channels and a tube length of \SI{60}{cm}. To achieve better impedance matching for the shorter tube length, the \SI{20}{\ohm} resistors on the readout interconnect board, as shown in Figure~\ref{sMDTandHedgehogCircuit}, were replaced with \SI{10}{\ohm} resistors. The chambers were operated with a \SI{1.2}{bar} He:iC\textsubscript{4}H\textsubscript{10} (90:10) gas mixture at an HV of \SI{1790}{V}, using a \SI{1}{GeV} hadron beam. During the measurement, one ATLAS sMDT FE board on each horizontal mini sMDT chamber was replaced with a preamplifier board under test, while the other six FE boards on the chambers remained in a working state and could provide reference tracks for further studies. The output signals from channels 0--15 of each preamplifier board were digitized by a CAEN DT5742 module~\cite{CAEN_DT5742}. Two stacked scintillators placed downstream of the mini sMDT chambers generated a coincidence signal that served as the trigger for the CAEN DT5742, enabling the selection of beam-induced events for performance evaluation of the preamplifier boards.
 
\begin{figure}[htbp]
\centering
\includegraphics[width=0.8\textwidth]{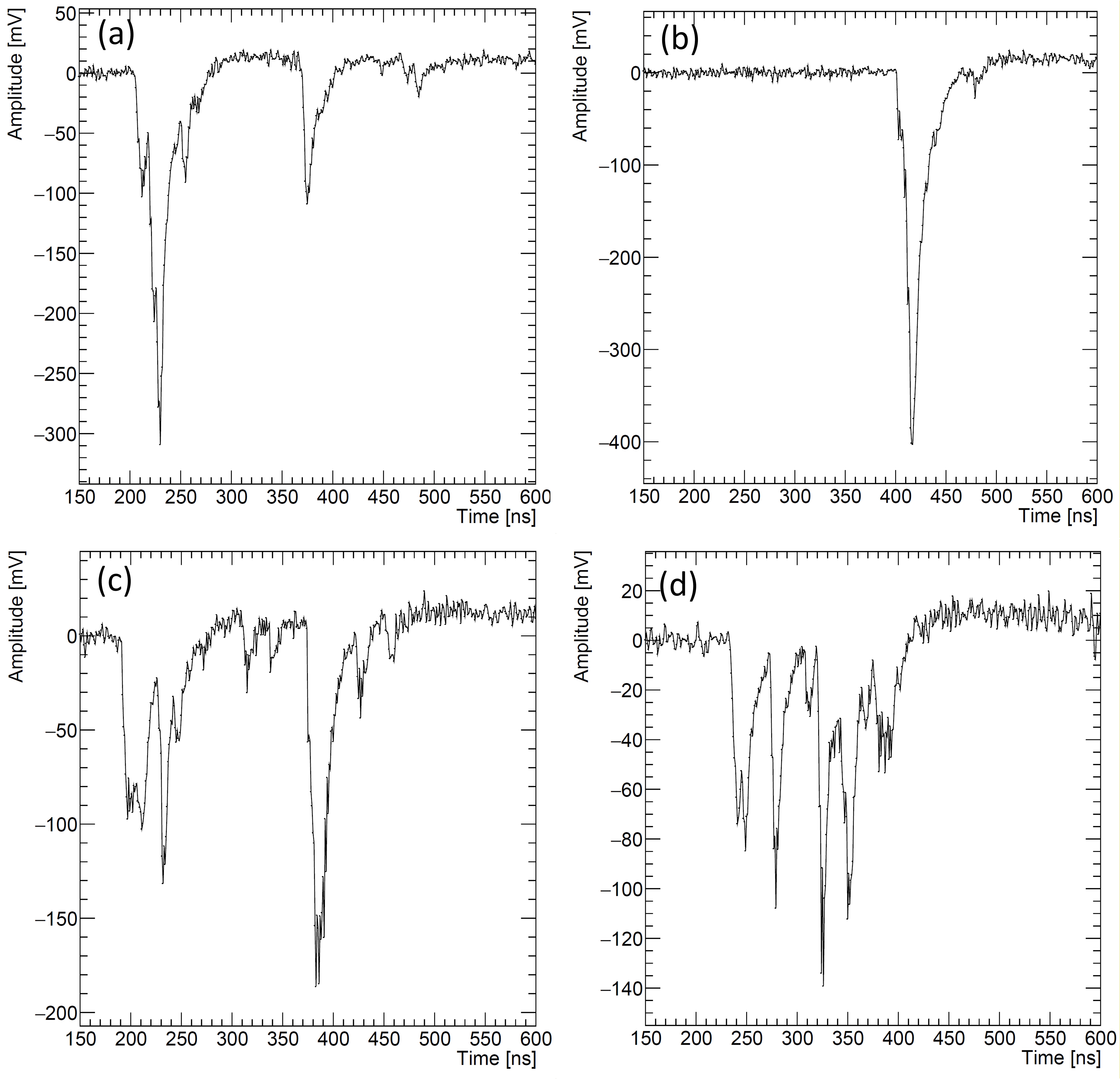}
\caption{Example output waveforms of the preamplifiers connected to the sMDT detectors in a \SI{1}{GeV} hadron beam, with the chamber operated at \SI{1790}{V} in \SI{1.2}{bar} He:iC\textsubscript{4}H\textsubscript{10} (90:10).}\label{waveform_detector}
\end{figure}

Figure~\ref{waveform_detector}  shows some representative output waveforms of the preamplifier produced by beam-induced signals. Multiple distinct peaks are visible, reflecting the relatively long ionization cluster intervals in the He:iC\textsubscript{4}H\textsubscript{10} (90:10) gas mixture at \SI{1.2}{bar}. The observed waveforms validate that the preamplifier achieves the signal quality necessary for reliable dN/dx measurements.

\begin{figure}[!ht]
    \centering
        \includegraphics[width=0.6\textwidth]{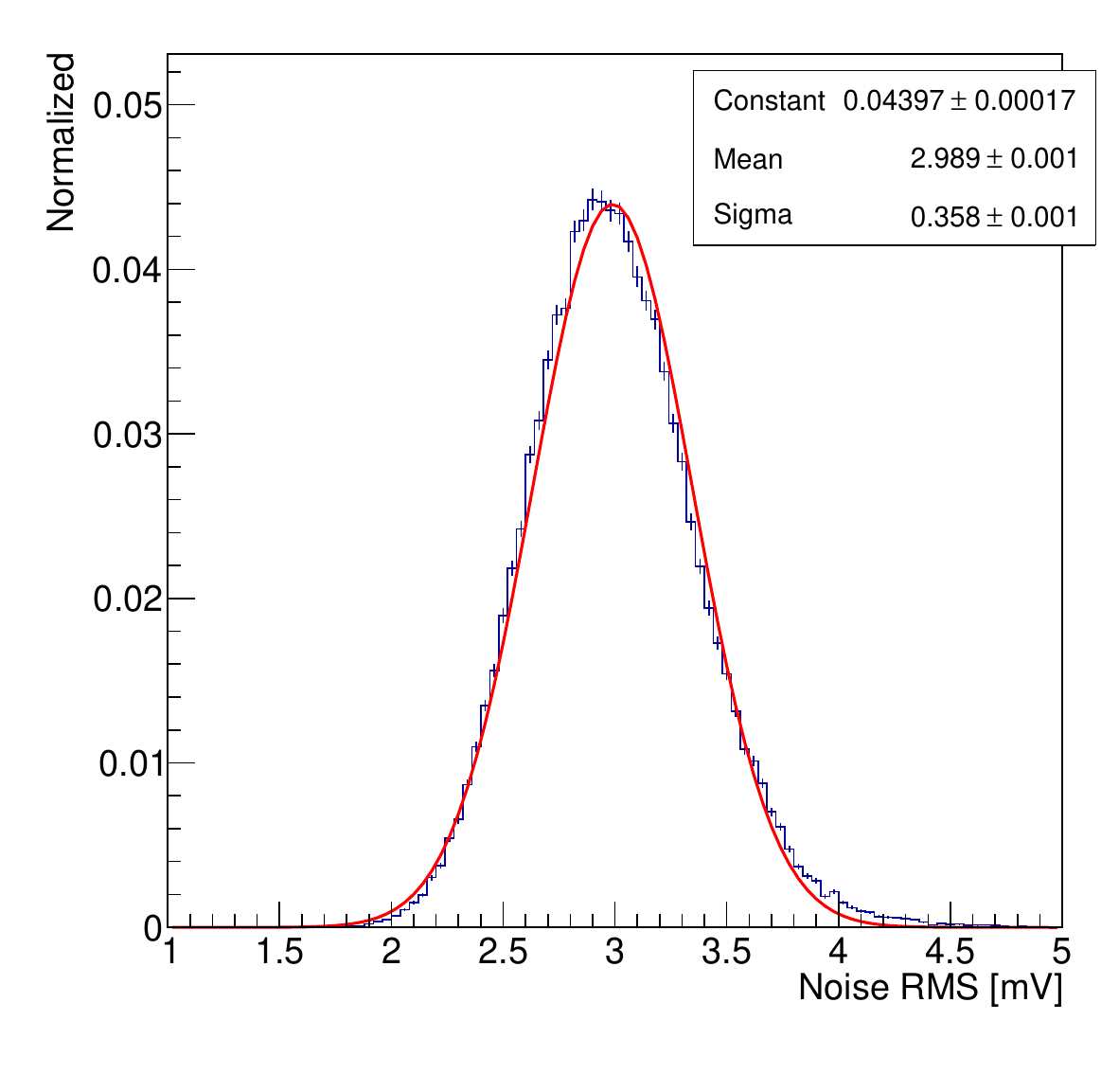}
    \caption{The distribution of the RMS noise from the preamplifier connected to the sMDT detector. "Constant","Mean", and "Sigma" are the fitting parameters of a Gaussian fit.}
    \label{fig:noise_rms}
\end{figure}

The RMS noise is extracted with the first 50 sampling points of the raw waveforms in one representative channel. Figure~\ref{fig:noise_rms} shows the distribution of the noise RMS for all raw waveforms, which is fitted with a Gaussian function. The mean value of the RMS noise obtained from the fit is  \( 2.99\,\text{mV}_{\text{rms}} \), corresponding to an ENC of approximately \( 0.14\,\text{fC} \) (\( 874\,\text{e}^{-} \)). As mentioned in Section~\ref {smdt_subsec2}, under a gas gain of \(10^5\), the charge of a single-electron cluster from the drift tube detector is approximately 0.8 fC. This yields an SNR of around 6 for a single-electron cluster, demonstrating that our amplifier can clearly resolve single-electron cluster signals. The noise of the preamplifier increases slightly upon installation on the sMDT detector. This increase is attributed to several factors: an added input capacitance (around  \SI{9}{\pF}) from the tube and readout interconnect board; the consequent thermal noise from their input resistance; and external interference coupled via the detector and interconnect board wiring. A Faraday cage is therefore essential to shield the preamplifier.

\begin{figure}[!ht]
    \centering
        \includegraphics[width=0.6\textwidth]{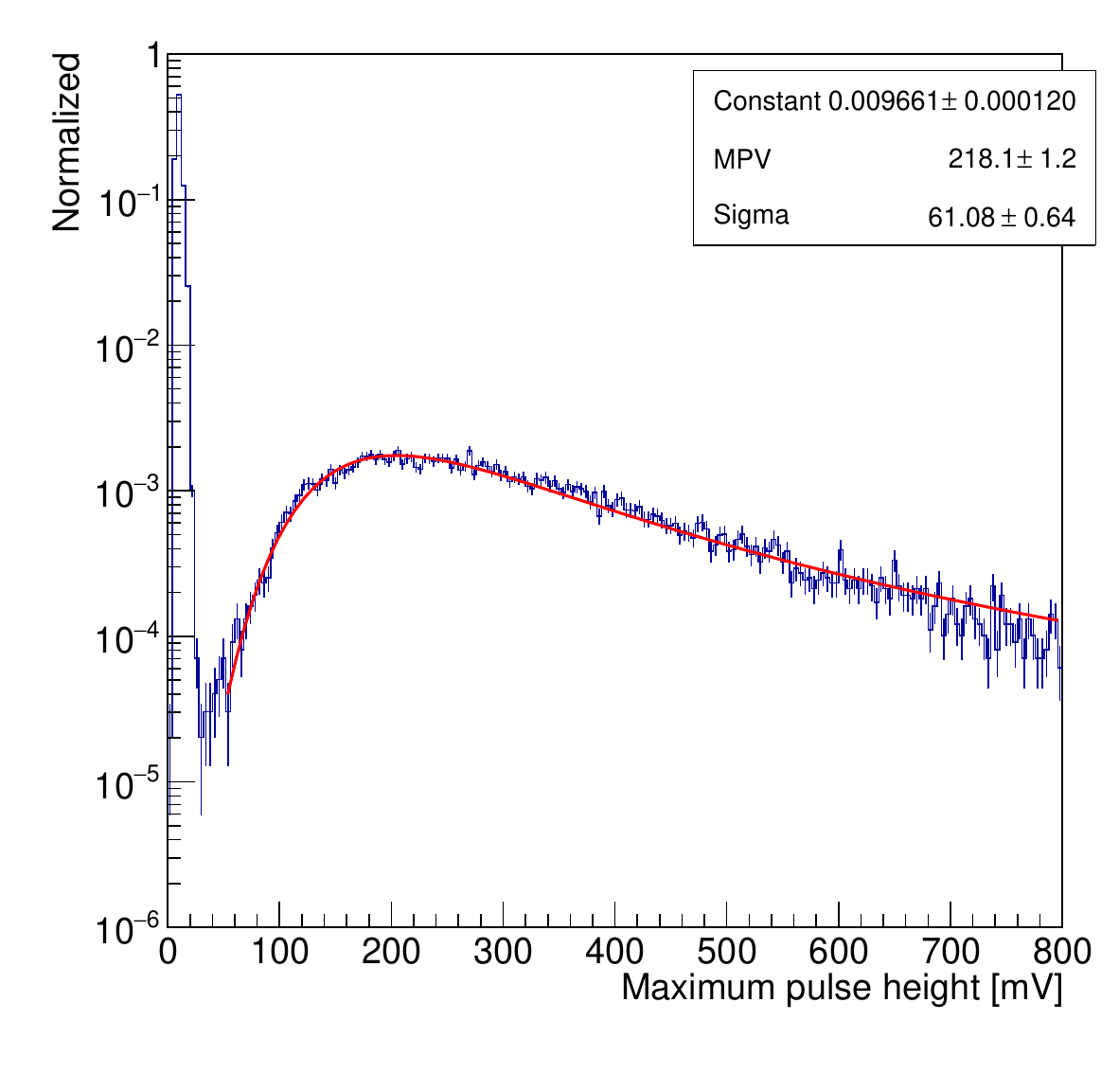}
    \caption{ Pulse amplitude distribution measured with the sMDT detector in a \SI{1}{GeV} hadron beam. The Landau fit yields an MPV of \SI{218.1}{mV}.}
    \label{fig_amplitude_distribution}
\end{figure}

Figure~\ref{fig_amplitude_distribution} shows the pulse amplitude distribution measured with the sMDT detector in a \SI{1}{GeV} hadron beam. The distribution is fitted with a Landau function, yielding a most probable value (MPV) of \SI{218.1}{mV}. With a measured RMS noise of \SI{2.99}{mV}, the SNR at the MPV is approximately 73. This high SNR confirms that the preamplifier provides adequate gain and low-noise performance for detecting minimum ionizing particles in a real drift tube detector.

\section{Conclusions}
\label{sec4}
A 24-channel ultra-low-noise preamplifier board has been designed to meet the requirements of drift tube detectors, which can be applied in the $dN/dx$ measurements for PID. The design employs a three-stage amplification topology, with the first two stages utilizing SiGe transistors to achieve very low noise and sufficient bandwidth for signal amplification. Some noise-minimization measures were implemented, resulting in outstanding overall performance. Electronic characterization shows a charge gain of \SI{21.11}{ \mV/\fC} with excellent linearity over an input dynamic range of \SIrange{0.3}{50}{\femto\coulomb}. The amplifier exhibits a $-3$\,dB bandwidth from approximately \SIrange{4}{542}{\mega\hertz} and a transition time of about \SI{640}{\pico\second}. Within the passband, the voltage gain reaches \SI{47.8}{dB}, accompanied by a voltage noise density of \SI{0.35}{\nano\volt\per\sqrt{\hertz}}. The preamplifier board achieves lower noise levels than most of the current state-of-the-art preamplifiers used in gaseous and silicon detectors. The preamplifier was validated using sMDT chambers operated with a \SI{1.2}{bar} He:iC\textsubscript{4}H\textsubscript{10} (90:10) gas mixture at \SI{1790}{V} in a test beam. The output signals showed multiple well-resolved clusters, with a measured integrated RMS noise of approximately \SI{2.99}{mV}. This corresponds to an ENC of \SI{0.14}{\femto\coulomb} and a SNR of 73, confirming the preamplifier's reliable performance under realistic drift-tube operating conditions. The successful results suggest that this amplifier architecture is also promising for broader applications in other gaseous or semiconductor detector systems.

\section*{CRediT authorship contribution statement}

    \textbf{Jiajin Ge}: Conceptualization, Methodology, Software, Writing – original draft;
    \textbf{Chihao Li}: Data curation, Writing – original draft;
    \textbf{Can Suslu}: Data curation, Writing – original draft;
    \textbf{Yuxiang Guo}: Methodology, Writing – review \& editing;
    \textbf{Emmett Salzer}: Resources, Writing – review \& editing;
    \textbf{Tiesheng Dai}: Methodology, Resources, Writing – review \& editing;
    \textbf{Jianming Qian}: Methodology, Resources, Writing – review \& editing;
    \textbf{Bing Zhou}: Supervision, Funding acquisition, Writing – review \& editing;
    \textbf{Junjie Zhu}: Project administration, Funding acquisition, Writing – original draft.

\section*{Declaration of Competing Interests}
\label{Acknowledgments}
The authors declare that they have no known competing financial interests or personal relationships that could have appeared to influence the work reported in this paper.

\section*{Acknowledgments}
\label{Acknowledgments}
We acknowledge the support of the U.S. Department of Energy under contracts DE-SC0007859 and DE-SC0012704. The authors would like to thank the PS/SPS Physics Coordination and the CERN PS beam line and infrastructure teams, and the DRD1 Collaborations.

\bibliographystyle{elsarticle_num}
\bibliography{reference}

\end{document}